\documentclass[11pt]{article}
\pdfoutput=1
\usepackage{amsmath}
\usepackage{amssymb}
\usepackage{amsthm}
\usepackage{graphicx}
\usepackage{amsfonts}
\usepackage{float}
\usepackage[margin=1in]{geometry}
\usepackage{lscape}
\usepackage{longtable}
\usepackage{url}
\usepackage[latin1]{inputenc}
\usepackage{tikz}
\usetikzlibrary{shapes,arrows}

\newtheorem{myprob}{Problem}
\newtheorem{thm}{Theorem}

\title{Five Quantum Algorithms Using Quipper}
\author{
Safat Siddiqui\\
\begin{small}Shahjalal University of Science and Technology\end{small}\\
\begin{small} Sylhet, Bangladesh\end{small}\\
\texttt{safat006@gmail.com}\\
\and
Mohammed Jahirul Islam\\
\begin{small}Shahjalal University of Science and Technology\end{small}\\
\begin{small} Sylhet, Bangladesh\end{small}\\
\texttt{jahir-cse@sust.edu}\\
\and
Omar Shehab\\ 
\begin{small}University of Maryland, Baltimore County, Maryland, USA\end{small}\\
\texttt{shehab1@umbc.edu}
}
\date{}

\edef\oldtt{\ttdefault}
\usepackage[scaled]{beramono}
\usepackage[T1]{fontenc}
\renewcommand*\ttdefault{\oldtt}
\newcommand{\bera}[1]{{\fontfamily{fvm}\selectfont #1}}
\begin{document}
\maketitle
\begin{small}
\begin{abstract}
Quipper is a recently released quantum programming language. In this report, we explore Quipper's programming framework by implementing the Deutsch's, Deutsch-Jozsa's, Simon's, Grover's, and Shor's factoring algorithms. It will help new quantum programmers in an instructive manner. We choose Quipper especially for its usability and scalability though it's an ongoing development project. We have also provided introductory concepts of Quipper and prerequisite backgrounds of the algorithms for readers' convenience. We also have written codes for oracles (black boxes or functions) for individual algorithms and tested some of them using the Quipper simulator to prove correctness and introduce the readers with the functionality. As Quipper 0.5 does not include more than \ensuremath{4 \times 4} matrix constructors for Unitary operators, we have also implemented \ensuremath{8 \times 8} and \ensuremath{16 \times 16} matrix constructors.
\end{abstract}

\section{Introduction}
Quantum computing is an interdisciplinary research area for physicists, mathematicians and computer scientists. Keeping pace with the developments in quantum hardware and algorithms, quantum programming languages are also developing. In this report\footnote{This report is based on the undergraduate thesis work of SS.}, we have proposed the implementations of five quantum algorithms, namely, Deutsch's algorithm, Deutsch-Jozsa's algorithm, Simon's periodicity algorithm, Grover's search algorithm and Shor's factoring algorithm using Quipper, a new functional quantum programming language. To our knowledge this report is the first such implementation of the above mentioned algorithms using Quipper. When we will have a physical gate-based quantum computer, these Quipper codes will guide us to get the real results from these algorithms instead of mere simulations. We choose these five algorithms because of their theoretical and pedagogical importance. We have also implemented oracles (black boxes or functions) to use them as inputs for those algorithms. As quantum hardwares are not available yet, we have used Quipper simulator to test some of them classically so that readers can test their own oracles. We assume readers have the initial knowledge of quantum data structure like qubit, quantum gates like \emph{Hadamard gate, controlled-not gate} etc. But it's not necessary to have previous experience of functional programming approach. We give some ideas about quantum programming languages in Section 2. In Section 3, we introduce Quipper and try to present a short tutorial. Implementations of the five quantum algorithms using Quipper are given in section 4. Here readers will get the basic ideas of quantum algorithms and also the basic structures of Quipper codes. Finally we draw our conclusion in section 5.


\section{Quantum programming languages}
Quantum programming language is an active area of quantum computational research. According to E. H. Knill's \cite{knill1996conventions} proposed architecture of quantum computers, the quantum machine has to be controlled by classical devices. Existing quantum programming languages are designed with classical controls such as loop, conditions etc and allow both quantum and classical data. Programming languages are mainly divided into two paradigms: \emph{imperative quantum programming languages} and \emph{functional quantum programming languages}. In imperative or procedural programming approach, programmers give instructions to the machines step by step, tell exactly how to do the task, and by functional or declarative programming approach, programmers tell the machines what to do, it is not necessary to tell how to do exactly. \emph{C, C++, Java, Python} etc are the examples of imperative programming approach and \emph{Scala, Erlang, Haskell} are the examples of functional programming approach. \emph{Quantum pseudocode} \cite{knill1996conventions}, \emph{QCL (Quantum Computing Language)} \cite{omer1998procedural}, \emph{Q language} \cite{bettelli2003toward}, \emph{qGCL (Quantum Guarded Command Language)} \cite{zuliani2001quantum} etc are imperative quantum programming languages and \emph{QFC} \cite{selinger2004towards}, \emph{QPL (Quantum Programming Language)} \cite{selinger2004towards}, \emph{cQPL (communication capable QPL)} \cite{mauerer2005semantics}, \emph{QML} \cite{altenkirch2005functional}, \emph{Quantum lambda calculi} \cite{van2004lambda}, \emph{Quipper} \cite{q1} etc are functional quantum programming languages \cite{w1}. These programming languages mainly use computational models like \emph{Quantum Turing Machine, Quantum Circuits, Quantum Lambda Calculus} etc. More details about quantum programming languages can be found in \cite{simonsurvey}, \cite{model}.

\section{What is Quipper?}
Quipper is an embedded functional quantum programming language \cite{q3}. This language is based on Haskell, a pure functional classical programming language. As Haskell's type system is one of the most powerful type systems, by using advanced features of it, Quipper provides many higher order and overloaded operators, though Haskell doesn't have linear type and dependent type features. To overcome this lacking, Quipper checks linear and dependent types in run time rather than in compile time. Thus Quipper offers a corrective, scalable and usable programming framework for quantum computation. As of 2014, Quipper is planning to be equipped soon with stand-alone compiler or at least a custom type-checker \cite{q1}.

Quipper was developed by Richard Eisenberg, Alexander S. Green, Peter LeFanu Lumsdaine, Keith Kim, Siun-Chuon Mau, Baranidharan Mohan, Won Ng, Joel Ravelomanantsoa-Ratsimihah, Neil J. Ross, Artur Scherer, Peter Selinger, Beno\^it Valiron, Alexandr Virodov and Stephan A. Zdancewic, in a research supported by the Intelligence Advanced Research Projects Activity (IARPA) \cite{IARPA}. It was first released in June 19, 2013 as a beta version 0.4 \cite{q3}.

\subsection{Quipper execution model}
Quipper program executes in three phases: \emph{compile time, circuit generation time} and \emph{circuit execution time}. \emph{Compile time} phase and \emph{circuit generation time} phase take place on classical computer. The last and final phase, \emph{Circuit execution time} occurs on physical quantum computer. 

In \emph{compile time} phase, Quipper takes source code, compile time parameters and uses Haskell's compiler to generate executable object code as output. \emph{Circuit generation time phase} takes the executable object code, circuit parameters (register size, problem size etc) as input and outputs a representation of quantum circuit. \emph{Circuit execution time} phase takes the quantum circuit, some circuit inputs (qubits fetched from long-term storage to initialize circuit inputs, if supported by quantum device, classical bits for classical circuit inputs).This phase outputs the measurement results of quantum subroutines in classical bits and moves the qubits (those are used as input) to long-term storage (if supported by quantum device).

\begin{figure}[H]

\begin{center}
\tikzstyle{decision} = [diamond, draw, fill=blue!20, 
    text width=4.5em, text badly centered, node distance=3cm, inner sep=0pt]
\tikzstyle{block} = [rectangle, draw, fill=blue!20, 
    text width=5em, text centered, rounded corners, minimum height=6em, minimum width = 5em]
\tikzstyle{line} = [draw, -latex', text width = 3 cm]
\tikzstyle{cloud} = [node distance=3cm,
    minimum height=2em]

\begin{tikzpicture}[ auto]
    \node [cloud] (start){};
    \node [block, right of = start, node distance = 3 cm] (first) {Compile time phase\\ in \\Classical device};
    \node [block, right of=first, node distance=5.2 cm] (second) {Circuit generation time phase\\ in \\Classical device};
    \node [block, below of=second, node distance=5.2 cm] (third) {Circuit execution time phase\\ in \\Quantum device};
    \node [cloud, right of = third, node distance =4.5 cm] (end){};
    \path [line] (start) -- node{source codes}(first);	
    \path [line] (first) -- node {executable object code} (second);
    \path [line] (second) -- node{ a representation of quantum circuit}(third);
    \path [line] (third) -- node{measurement results in classical bits} (end);
\end{tikzpicture}
\caption{Quipper execution model}

\end{center}

\end{figure}
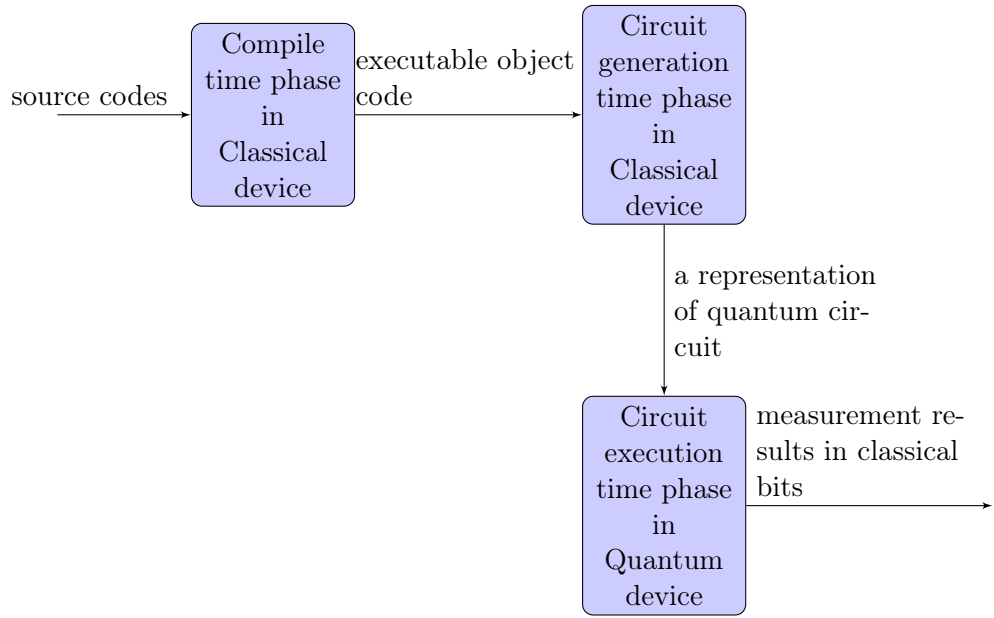 
In the model of Quipper execution, classical controller generates a circuit according to the source codes, sends it to quantum hardware for execution and takes the measurement results. Quipper provides \bera{print\_generic} and \bera{print\_simple} functions to print the circuit in available output format (such as text, PostScript, and PDF). Quipper also provides \bera{run\_generic} function to simulate the circuit on classical machine. More details about Quipper can be found in \cite{q1}.

\subsection{Quipper examples}

Quipper has three basic data types, \bera{Bit, Bool} and \bera{Qubit}. \bera{Bit, Bool} represent classical data and \bera{Qubit} is for quantum data. Quipper distinguishes between \emph{parameter} and \emph{input}. When the value is known at \emph{circuit generation time} phase it is \emph{parameter} and when the value is known only at \emph{circuit execution time} phase it is called \emph{input}. Here \bera{Bool} is a boolean \emph{parameter}, \bera{Bit} and \bera{Qubit} are respectively a classical boolean \emph{input} and a quantum \emph{input} to a circuit. \bera{Bool} can be converted to \bera{Bit}, but vice versa is not possible. Quantum measurements are \bera{Bit}s rather than \bera{Bool}s, as measurements occur at \emph{circuit execution time} phase. Some data has both \emph{parameter} and \emph{input} components, this type of data is called \emph{shape}. Circuit size is a \emph{shape} type data, here qubit \emph{list} (data structure) is \emph{input} type and the length of the \emph{list} is \emph{parameter} type.

In this section, we will write our first Quipper code, see how to apply quantum gates, be familiar with the built-in data structures and finally, know the measurement operation that Quipper provides.
\paragraph{The \emph{Hello World} program :} To start with Quipper we will write a simple \emph{hello world} function. This function will take a classical data \bera{Bool} and return a corresponding quantum data \bera{Qubit}. The code is given below: \\

\begin{footnotesize}
\bera{
import Quipper\\

-\-- declare hello\_world function

hello\_world :: Bool \ensuremath{\rightarrow} Circ Qubit

hello\_world var = do

~~-\-- convert \bera{Bool} into \bera{Qubit}

~~qbit \ensuremath{\leftarrow} qinit var

~~-\-- to label a variable on pdf circuit

~~label (qbit) ("\ensuremath{|1\rangle}")

~~-\-- return the result

~~return qbit
\\}
\end{footnotesize}

Let us focus on the important parts: first we import the \bera{Quipper} library to get all Quipper properties, built-in functions, operators, data types etc. The first line of a function is the type signature of that function. Here the type signature means \bera{hello\_world} is a function that takes a \bera{Bool} type data and returns a \bera{Qubit} type data. Arguments are separated by "\ensuremath{\rightarrow}" notation and \bera{Circ} is a type operator (in Haskell, it is called \bera{Monad}) that represents this function can have a side effect when it is evaluated. Usually functions are written in a \bera{do} block. A \bera{do} block starts with the \bera{do} keyword and then followed by a series of expressions or operations. The variable \bera{var} will store the value that will be passed through this function. In Quipper "-\-- ..." and "\{- ... -\}" are used for commenting codes. The \bera{qinit} operator (from \bera{Quipper} library) takes a \bera{Bool} as input and initializes a \bera{Qubit}, a quantum state corresponding to the classical data \bera{Bool} (If \bera{False} then qubit is \ensuremath{|0\rangle} and if \bera{True} then \ensuremath{|1\rangle}). "\ensuremath{\leftarrow}" notation means that the new quantum state is stored in a variable \bera{qbit} and finally returns the state.

The starting point of a Quipper program is the \bera{main} function. As we have mentioned before, \emph{circuit generation time} phase sends a circuit of the source code to the physical quantum device to execute and \bera{print\_generic} and \bera{print\_simple} functions are used to print the circuit in an available output format. \bera{print\_simple} is used when circuit \emph{shape} is fixed and \bera{print\_generic} is used for the circuit which \emph{shape} is not fixed. Quipper provides \bera{label, comment, comment\_with\_label} for commenting on a generated circuit in a pdf document. In the \bera{main} function, using \bera{print\_simple} we will call the \bera{hello\_world} function with a \bera{Bool} type data \bera{True} and get a corresponding quantum state. After successful compilation, when we will execute this program, it generates a circuit in a pdf document. Circuits are read from left to right and wires are represented by horizontal lines. Quipper uses "\includegraphics{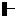}" notation to denote the allocation of a new qubit for corresponding classical data. The \bera{main} function is:\\

\begin{footnotesize}
\bera{main = print\_simple Preview (hello\_world True)}
\end{footnotesize}

\begin{figure}[H]
\centering
\includegraphics[width=.25\textwidth]{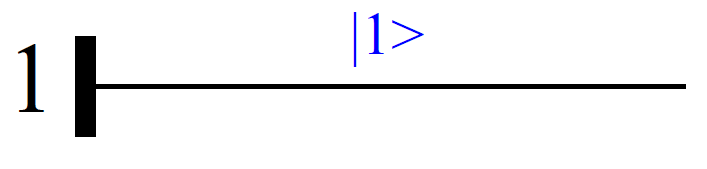}
\caption{Circuit for \bera{hello\_world} function}
\end{figure}

\paragraph{Apply Quantum Gates :} From \bera{hello\_world} function, we get a quantum state. Now we will apply a quantum gate on that qubit. At this point, we will take the advantage of Quipper's higher order functionality that means a function can be passed through another function. We will name our function \bera{apply\_gate} and pass \bera{hello\_world} function as parameter. \bera{apply\_gate} function will use \bera{hello\_world} to convert a classical state into a quantum state and apply a quantum gate on it. The code is given below:\\ 

\begin{footnotesize}
\bera{
-\-- declare apply\_gate function

apply\_gate :: (Bool \ensuremath{\rightarrow} Circ Qubit) \ensuremath{\rightarrow} Bool \ensuremath{\rightarrow} Circ Qubit

apply\_gate func bool = do

~~-\-- use func function to get qubit

~~qbit \ensuremath{\leftarrow} func bool

~~-\-- to comment on pdf circuit

~~comment "before gate"

~~-\-- apply \emph{Hadamard Transfomation}

~~qbit \ensuremath{\leftarrow} hadamard qbit

~~-\-- both label and comment

~~comment\_with\_label "after gate" (qbit) ("\ensuremath{(|0\rangle - |1\rangle)/2}")

~~-\-- return result

~~return qbit\\

-\-- main function to call the whole program

main = print\_simple Preview (apply\_gate hello\_world True)
\\}
\end{footnotesize}

\begin{figure}[H]
\centering
\includegraphics[scale=0.4]{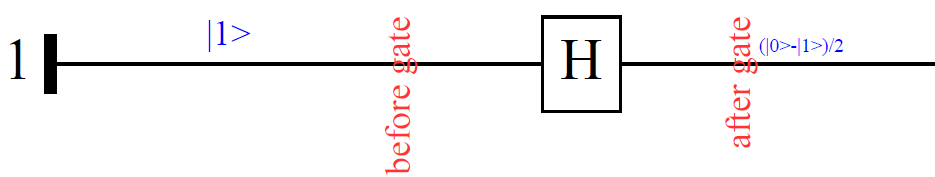}
\caption{Circuit for \bera{apply\_gate} function}
\end{figure}

The "type signature" of \bera{apply\_gate} tells that it has two arguments. First one is a function (Bool \ensuremath{\rightarrow} Circ Qubit) that takes a \bera{Bool} and returns a \bera{Qubit}, second one is a \bera{Bool} data type. This function returns a \bera{Qubit}. To apply \emph{Hadamard Transformation} we can use \bera{hadamard} or \bera{gate\_H} operator (box represents quantum gate in pdf of Quipper). We can also use \bera{gate\_H\_at} or \bera{hadamard\_at}, these operators don't return any value, in that case expression should be like \bera{hadamard\_at qbit}. We may use other quantum gates like \bera{qnot, gate\_X, gate\_S, gate\_Z, gate\_T, gate\_Y} etc.

In the \bera{main} function, we pass the \bera{hello\_world} function and a \bera{"True" Bool} type data through \bera{apply\_gate} function. \bera{hello\_world} and \bera{True} values are stored respectively in \bera{func} and \bera{bool} variables. Finally this function returns a quantum state after applying \emph{hadamard transformation}.\\

\paragraph{Data Structures:}
As Quipper's host language is Haskell, it mainly uses Haskell's data structures. Haskell provides many data structures like \emph{Map, Set} etc, but now we would like to focus on the most basic data structures that are \emph{list} and \emph{tuple}. \bera{let} and \bera{where} clauses are used for local bindings (we will use these in the implementations).

\emph{list} in Haskell is a linked list, it uses \bera{(:)} operator to bind an element. \bera{(++)} operator is used for concatenation. \bera{head} operator is used to get the first element of a \emph{list} and \bera{tail} is used to get the rest. Operator \bera{(!!)} is used to find the element of an index. \bera{[Bool]} and \bera{[Qubit]} are the example of \emph{list}. \emph{list}s are \emph{homogeneous} that means a single \emph{list} can contain only single type of elements. On the other hand, \emph{tuple} is a fixed number of single or different type components. \emph{tuple} is mainly used for returning multiple values of different data types. \bera{(Qubit, Bool)} and \bera{([Qubit], [Qubit])} are the examples of \emph{tuple}.

Previously we performed \emph{hadamard transformation} only on one qubit. To perform \emph{hadamard transformation}s on multiple qubits, we will use \emph{list} of qubits, \bera{[Qubit]}. In that case, we will need to modify the type signature of \bera{hello\_world} and \bera{apply\_gate} functions like \bera{hello\_world :: [Bool] \ensuremath{\rightarrow} Circ [Qubit]} and \bera{apply\_gate :: ([Bool] \ensuremath{\rightarrow} Circ [Qubit]) \ensuremath{\rightarrow} [Bool] \ensuremath{\rightarrow} Circ [Qubit]}. To perform quantum gates on a \emph{list}, we will use \bera{mapUnary} operator. Then expression should be like \bera{mapUnary hadamard qbit} instead of \bera{qbit \ensuremath{\leftarrow} hadamard qbit}, in \bera{apply\_gate} function.

In the \bera{main} function we will use \bera{replicate} operator and \bera{where} clause to create \bera{n True Bool} type data. We will also use \bera{print\_generic} as circuit \emph{shape} is not fixed here. The \bera{main} function is:\\

%
%
%
%
%
%
%
%
%
%
%
%
%
%
%
%
%
%
%
%
%
%
%
%
%
%

\begin{footnotesize}
\bera{
-\-- main function to call the whole program

main = print\_generic Preview (apply\_gate hello\_world (replicate n True))

~~where

~~~~n = 5
}
\end{footnotesize}

\begin{figure}[H]
\centering
\includegraphics[scale=0.4]{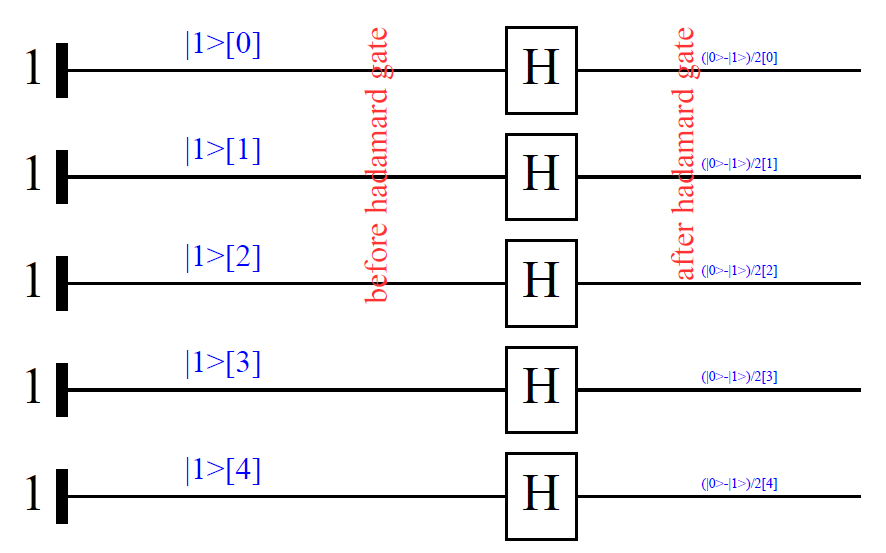}
\caption{Circuit of applying gates on multiple qubits}
\end{figure}

\paragraph{Measurement Operation :} Quipper provides \bera{measure} operator to measure quantum states and \bera{cdiscard} operator to discard classical data. These two are \emph{generic operator}s that means any data structure can be applied here. \bera{measure} operator takes \bera{Qubit} and collapses it to one of the basic states. \bera{cdiscard} operator takes classical \bera{Bit}s and discards. \includegraphics[width=0.05\textwidth]{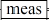} and \includegraphics[width=0.02\textwidth]{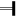} notations are used respectively to denote \bera{measure} and \bera{cdiscard} operations. We will use these operators in our \bera{controlled\_gate} function that applies a controlled gate operation on two qubits and measures their values. Corresponding code is given below:\\

\begin{footnotesize}
\bera{
-\-- declare controlled\_gate function

controlled\_gate :: (Bool, Bool) \ensuremath{\rightarrow} Circ Bit

controlled\_gate (cntrl\_qbit, trget\_qbit) = do

~~-\-- convert Bool into Qubit

~~cntrl\_qbit \ensuremath{\leftarrow} qinit cntrl\_qbit

~~trget\_qbit \ensuremath{\leftarrow} qinit trget\_qbit

~~-\-- controlled gate operation

~~gate\_X\_at trget\_qbit `controlled` cntrl\_qbit

~~-\-- measure Qubits

~~(cntrl\_qbit, trget\_qbit) \ensuremath{\leftarrow} measure (cntrl\_qbit, trget\_qbit)

~~-\-- discard value

~~cdiscard trget\_qbit

~~-- return result

~~return cntrl\_qbit\\

-\-- main function

main = print\_simple Preview (controlled\_gate (False,True))
\\}
\end{footnotesize}

\begin{figure}[H]
\centering
\includegraphics[width=.25\textwidth]{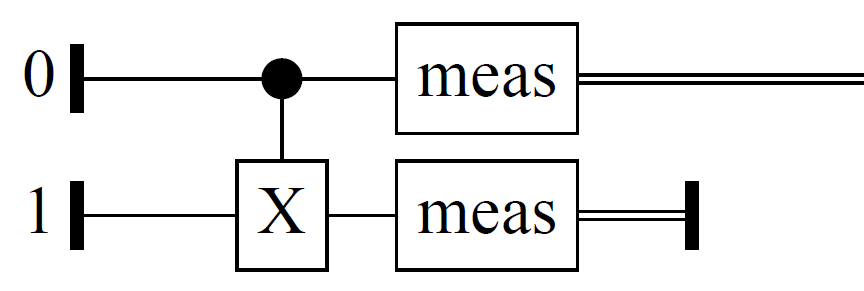}
\caption{Circuit of measurement example}
\end{figure}
Here we apply a \bera{X} gate on \bera{trget\_qbit} with the control of \bera{cntrl\_qbit}. To do that we use \bera{`controlled`} operator. We can also use this operator for \emph{list} data structure. Then the expression will be like \bera{`controlled` list}. To specify their controlling states we can write the expression like \bera{`controlled` list .==. [0,1,0]} (assume \bera{list} has three elements).

Quipper gives the opportunity to apply "Unitary Operations" using matrix. We can do the previous example using \ensuremath{4 \times 4} matrix. We will import some additional libraries and write a function \bera{operator} that have the matrix.\\

\begin{footnotesize}
\bera{

-\-- import Matrix constructors

import Libraries.Synthesis.Matrix

import QuipperLib.Synthesis

import Libraries.Synthesis.Ring\\

-\-- initialize unitary matrix

operator :: Matrix Four Four DOmega

operator = matrix4x4 ( 1, 0, 0, 0 )

~~~~~~~~~~~~~~~~~~~~~( 0, 1, 0, 0 )

~~~~~~~~~~~~~~~~~~~~~( 0, 0, 0, 1 )

~~~~~~~~~~~~~~~~~~~~~( 0, 0, 1, 0 ) 
\\}
\end{footnotesize}

In \bera{controlled\_gate} function, if we change \bera{gate\_X\_at trget\_qbit `controlled` cntrl\_qbit} expression into \bera{exact\_synthesis operator [cntrl\_qbit, trget\_qbit]} expression, \bera{main} function will generate the same previous circuit. More tutorials on Quipper can be found in \cite{q2}.

\section{Implementation of quantum algorithms}
Quantum algorithms are interesting because for some cases it gives exponential computational power like factoring integers, finding orders of functions. Instead of being one state at a time, quantum computer gives the opportunity to the states to be in a superposition. To get the superposition, many quantum algorithms initialize \ensuremath{n} qubits with \ensuremath{|0\rangle} and apply \emph{Hadamard transformation}. It maps \ensuremath{n} qubits to the superpositions of all \ensuremath{2^n} orthogonal states in the \ensuremath{|0\rangle, |1\rangle} basis with equal weight.

Here we implement five quantum algorithms in Quipper Programming Language. Section 4.1 describes the Deutsch's algorithm which determines the fairness of a boolean function. Section 4.2 is the Deutsch-Jozsa algorithm which is the generalized version of Deutsch's algorithm. Section 4.3 is Simon's periodicity algorithm that finds the hidden pattern of a function. Section 4.4 describes Grover's search algorithm that can search an element from unordered array in \ensuremath{\sqrt{n}} time. Finally section 4.5 describes Shor's factoring algorithm which can factor integers in polynomial time.

\subsection{Deutsch algorithm}
The Deutsch algorithm, published in 1985 \cite{deutsch1985quantum}, solves a contrived problem to see how quantum computers can be used. This algorithm actually determines if a function is one to one or not. A function is called \emph{balanced} if \ensuremath{f(0)\neq f(1)}, means one to one. Otherwise a function is called \emph{constant} if \ensuremath{f(0)=f(1)}. Deutsch's algorithm solves the following problem (we quote the definition of the problem verbatim from \cite{noson}):

\begin{myprob}Given a function \ensuremath{f : \{0,1\}\rightarrow \{0,1\}} as a black box, where one can evaluate an input, but cannot "look inside" and "see" how the function is defined, determine if the function is balanced or constant.\end{myprob}

Classical algorithm needs two steps (first step to find \ensuremath{f(0)}'s value and second step to find \ensuremath{f(1)}'s value) to determine and Deutsch algorithm needs one step. This algorithm provides an oracle separation between P and EQP.

First, we will write codes for Deutsch algorithm. Then we will code oracles (black boxes) of balanced and constant functions and finally, test those oracles using the simulator provided by Quipper. 

\subsubsection{Quantum circuit for Deutsch algorithm}
As we have to test an oracle whether it is balanced or constant, at first we will write our own data type named \bera{Oracle}. Later we will declare our oracles of functions using this data type. Main section of Deutsch algorithm will be in \bera{deutsch\_circuit} function. Finally \bera{main} function will call \bera{deutsch\_circuit} function with an oracle named \bera{empty\_oracle}. here \bera{empty\_oracle} is a dummy oracle, we will use this to generalize the circuit. Later we will use some working oracles. \bera{main} function will get a classical data \bera{Bit} to determine whether given oracle is \emph{balanced} or \emph{constant}. The code for \bera{deutsch\_circuit} is:\\

\begin{footnotesize}
\bera{
import Quipper\\

-\-- declare Oracle data type

data Oracle = Oracle\{

~~-\-- oracle of function \ensuremath{f(x)}

~~oracle\_function :: (Qubit,Qubit) \ensuremath{\rightarrow} Circ (Qubit,Qubit)

\}\\

-\-- declare deutsch\_circuit function

deutsch\_circuit :: Oracle \ensuremath{\rightarrow} Circ Bit

deutsch\_circuit oracle = do

~~-\-- create the ancillae

~~top\_qubit \ensuremath{\leftarrow} qinit False

~~bottom\_qubit \ensuremath{\leftarrow} qinit True

~~label (top\_qubit, bottom\_qubit) ("\ensuremath{|0\rangle}","\ensuremath{|1\rangle}")

~~-\-- do the first Hadamards

~~hadamard\_at top\_qubit

~~hadamard\_at bottom\_qubit

~~comment "before oracle"

~~-\-- call the oracle

~~oracle\_function oracle (top\_qubit, bottom\_qubit)

~~comment "after oracle"

~~-\-- do the last Hadamards

~~hadamard\_at top\_qubit

~~-\-- measure qubits 

~~(top\_qubit, bottom\_qubit) \ensuremath{\leftarrow} measure (top\_qubit, bottom\_qubit)

~~-\-- discard un-necessary output and return the result

~~cdiscard bottom\_qubit

~~return top\_qubit\\

-\-- main function to call the whole program

main = print\_generic Preview (deutsch\_circuit empty\_oracle)

~~~where

~~~~~-\-- declare empty\_oracle's data type

~~~~~empty\_oracle :: Oracle

~~~~~empty\_oracle = Oracle \{ 

~~~~~~oracle\_function = empty\_oracle\_function

~~~~~\}

~~~~~-\-- initialize empty\_oracle

~~~~~empty\_oracle\_function:: (Qubit,Qubit) \ensuremath{\rightarrow} Circ (Qubit,Qubit)

~~~~~empty\_oracle\_function (one,two) = named\_gate "Oracle" (one,two)
}\\
\end{footnotesize}

\begin{figure}[H]
\centering
\includegraphics[scale=0.4]{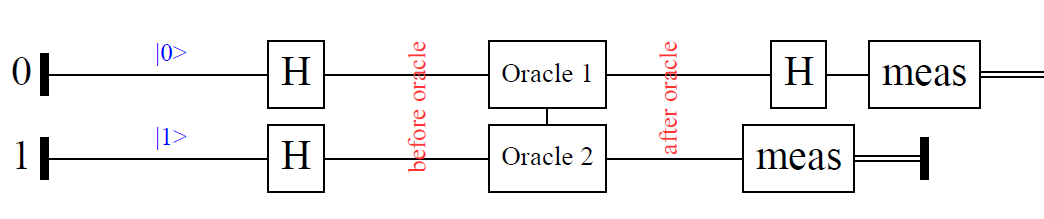}
\caption{Circuit for Deutsch algorithm \cite{noson}}
\end{figure}

\subsubsection{Oracle examples}
\paragraph{Balanced Oracle :}
We will write code for balanced oracle in \bera{balanced\_oracle}. Here we will perform a controlled not operation. This will be a balanced oracle because here \ensuremath{f(0) \neq f(1)}. Finally \bera{main} function will pass this oracle to \bera{deutsch\_circuit} function. So the modified section is:\\

\begin{footnotesize}
\bera{
main = print\_generic Preview (deutsch\_circuit balanced\_oracle)

~~~where

~~~~~-\-- declare balanced\_oracle's data type

~~~~~balanced\_oracle :: Oracle

~~~~~balanced\_oracle = Oracle \{ 

~~~~~~oracle\_function = balanced\_oracle\_function

~~~~~\}

~~~~~-\-- initialize oracle function \ensuremath{f(x)}

~~~~~balanced\_oracle\_function:: (Qubit,Qubit) \ensuremath{\rightarrow} Circ (Qubit,Qubit)

~~~~~balanced\_oracle\_function (x,y) = do

~~~~~~qnot\_at y `controlled` x

~~~~~~return (x,y)
\\}
\end{footnotesize}

\begin{figure}[H]
\centering
\includegraphics[scale=0.4]{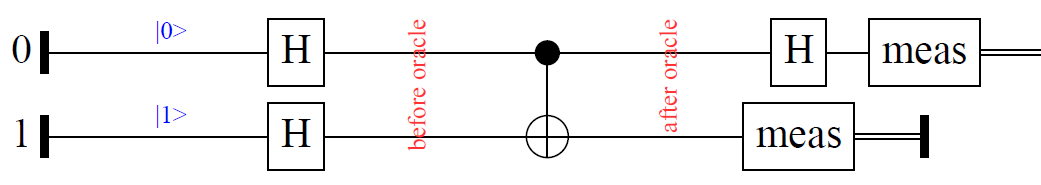}
\caption{Circuit for balanced oracle}
\end{figure}

\paragraph{Constant Oracle:}
We will write code for constant oracle in \bera{constant\_oracle}. We will remain qubits states same so that this oracle will ensure \ensuremath{f(0)=f(1)}. In \bera{main} function, we will pass this oracle to \bera{deutsch\_circuit} function. So the modified section is:\\

\begin{footnotesize}
\bera{
main = print\_generic Preview (deutsch\_circuit constant\_oracle)

~~~where

~~~~~-\-- declare constant\_oracle's data type

~~~~~constant\_oracle :: Oracle

~~~~~constant\_oracle = Oracle \{ 

~~~~~~oracle\_function = constant\_oracle\_function

~~~~~\}

~~~~~-\-- initialize oracle function \ensuremath{f(x)}

~~~~~constant\_oracle\_function:: (Qubit,Qubit) \ensuremath{\rightarrow} Circ (Qubit,Qubit)

~~~~~constant\_oracle\_function (x,y) = do

~~~~~~-\-- Qubits will remain the same

~~~~~~return (x,y)
\\}
\end{footnotesize}

\begin{figure}[H]
\centering
\includegraphics[scale=0.4]{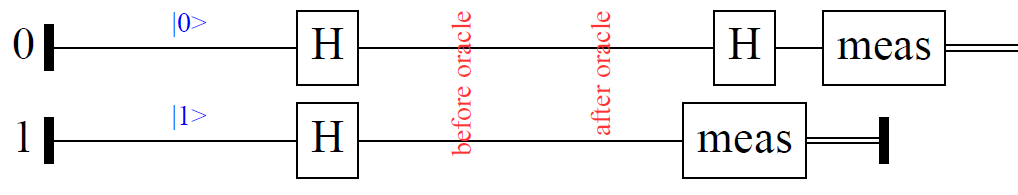}
\caption{Circuit for constant oracle}
\end{figure}

\subsubsection{Simulation}
We can test previous oracles with the simulator included in Quipper. We will remove \bera{balanced\_oracle}, \bera{constant\_oracle} functions from \bera{where} clause and write those functions independently. We will write two new functions \bera{simulate} and \bera{circuit}. In \bera{simulate} function, we will test the oracle using built-in \bera{run\_generic} function and in \bera{circuit} function, we will show the simulation results. \bera{main} function is to start the whole program.\\

%
%
%
%
%
%
%
%
%
%
%
%
%
%
%
%
%
%
%
%
\begin{footnotesize}
\bera{
-\-- import modules for simulations

import qualified Data.Map as Map

import QuipperLib.Simulation

import System.Random\\

-\-- declare simulate function

simulate :: Circ Bit \ensuremath{\rightarrow} Bool

simulate oracle = (run\_generic (mkStdGen 1) (1.0::Double) oracle)
\\}
\end{footnotesize}

Let's highlight Quipper's built-in \bera{run\_generic} function for simulation. It takes three arguments:

\begin{enumerate}
\item A source of randomness, something of type \bera{StdGen} from Haskell's \bera{System.Random} library. \bera{mkStdGen} is a function creating such an object out of an \bera{Int}. So \bera{(mkStdGen~1)} does the trick.
\item An instance of real number so that the function \bera{run\_generic} knows what to use as datatype for reals. Here we take \bera{Double}.
\item The circuit to run.
\end{enumerate}

\bera{circuit} function will take \bera{simulate} function and an \bera{Oracle} . \bera{IO()} represents this function has an I/O operation. \bera{simulate} function will return a \bera{Bool} type data \bera{True} when oracle is balanced and \bera{False} when oracle is constant. \bera{circuit} function will take the result and perform an I/O operation. \bera{main} function should be modified. When we will run the program in command prompt, we get the simulation results "Given oracle is Balanced" for balanced oracles and "Given oracle is Constant" for constant oracles.\\

\begin{footnotesize}
\bera{
-\-- declare circuit function

circuit :: (Circ Bit \ensuremath{\rightarrow} Bool) \ensuremath{\rightarrow} Oracle \ensuremath{\rightarrow} IO ()

circuit run oracle =

~~-\-- first deutsch\_circuit will apply on oracle

~~-\-- then run function will evaluate the result

~~if run (deutsch\_circuit oracle) 

~~then putStrLn "Given oracle is Balanced"

~~else putStrLn "Given oracle is Constant"\\

-\-- main function

main = do

~~-\-- test constant\_oracle

~~circuit simulate constant\_oracle

~~-\-- test balanced\_oracle

~~circuit simulate balanced\_oracle
\\}
\end{footnotesize}

\begin{figure}[H]
\centering
\includegraphics[width=.60\textwidth]{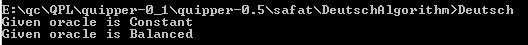}
\caption{Simulation of Deutsch Algorithm}
\end{figure}

\subsection{Deutsch-Jozsa algorithm}
The Deutsch-Jozsa algorithm, proposed by David Deutsch and Richard Jozsa in 1992 \cite{deutsch1992rapid}, is the generalized version of Deutsch algorithm, which accepts a string of \ensuremath{n~0}'s and \ensuremath{1}'s and outputs a zero or one.

A function is called \emph{balanced} if exactly half of the input's outputs are \ensuremath{0}'s and other half of the input's outputs are \ensuremath{1}'s. And a function is called \emph{constant} if all the input's outputs are either \ensuremath{0}'s or \ensuremath{1}'s. Deutsch-Jozsa algorithm solves the following problem (we quote the definition of the problem verbatim from \cite{noson}):

\begin{myprob}Given a function \ensuremath{f : \{0,1\}^n\rightarrow \{0,1\}} as a black box, where one can evaluate an input, but cannot "look inside" and "see" how the function is defined, determine if the function is balanced or constant.\end{myprob} 

For classical algorithm, the best case scenario is when first two inputs have different outputs which ensures that given function is balanced. But to ensure a function is constant, it must evaluate the function more than half of the possible inputs. So it requires \ensuremath{\frac{2^n}{2}+1 = 2^{(n-1)}+1} evaluations. But Deutsch-Jozsa algorithm solves this problem in one evaluation, provides an oracle separation of P and EQP, that's an exponential speedup.

\subsubsection{Circuit of Deutsch-Jozsa algorithm}
Unlike Deutsch algorithm, the Deutsch-Jozsa algorithm accepts a string of length \ensuremath{n}. That's why we will need to augment our previous \bera{Oracle} data type so that it can deal with the string of qubits. The core section of Deutsch-Jozsa algorithm will be in \bera{deutsch\_jozsa\_circuit} function (more or less similar to \bera{deutsch\_circuit} function). In the \bera{main} function, we will call \bera{deutsch\_jozsa\_circuit} with a dummy oracle named \bera{empty\_oracle}. So the code for Deutsch-Jozsa algorithm is:\\

\begin{footnotesize}
\bera{
import Quipper\\

-\-- declare modified Oracle data type

data Oracle = Oracle \{    
 
~~~-\-- declare the length of a string
 
~~~qubit\_num :: Int,

~~~-\-- declare oracle function \ensuremath{f(x)}

~~~function :: ([Qubit], Qubit) \ensuremath{\rightarrow} Circ ([Qubit], Qubit)  

\}\\

-\-- declare deutsch\_jozsa\_circuit function

deutsch\_jozsa\_circuit :: Oracle \ensuremath{\rightarrow} Circ [Bit]

deutsch\_jozsa\_circuit oracle = do

~~-\-- initialize string of qubits

~~top\_qubits \ensuremath{\leftarrow} qinit (replicate (qubit\_num oracle) False)  

~~bottom\_qubit \ensuremath{\leftarrow} qinit True

~~label (top\_qubit, bottom\_qubit) ("\ensuremath{|0\rangle}","\ensuremath{|1\rangle}")

~~-\-- do the first hadamard

~~mapUnary hadamard top\_qubits

~~hadamard\_at bottom\_qubit

~~comment "before oracle"

~~-\-- call oracle

~~function oracle (top\_qubits, bottom\_qubit)

~~comment "after oracle"

~~-\-- do the last hadamard

~~mapUnary hadamard top\_qubits

~~-\-- measure qubits

~~(top\_qubits, bottom\_qubit) \ensuremath{\leftarrow} measure (top\_qubits, bottom\_qubit)

~~-\-- discard unnecessary output and return result

~~cdiscard bottom\_qubit

~~return top\_qubits\\

-\-- main function

main = print\_generic Preview (deutsch\_jozsa\_circuit empty\_oracle) 

~~where 

~~~~-\-- declare empty\_oracle's data type

~~~~empty\_oracle :: Oracle

~~~~empty\_oracle = Oracle \{
 
~~~~~qubit\_num = 5,

~~~~~function = empty\_oracle\_function

~~~~\}

~~~~-\-- initialize empty\_oracle's function \ensuremath{f(x)}

~~~~empty\_oracle\_function:: ([Qubit],Qubit) \ensuremath{\rightarrow} Circ ([Qubit],Qubit)

~~~~empty\_oracle\_function (ins,out) = named\_gate "Oracle" (ins,out)
\\}
\end{footnotesize}

\begin{figure}[H]
\centering
\includegraphics[scale=0.4]{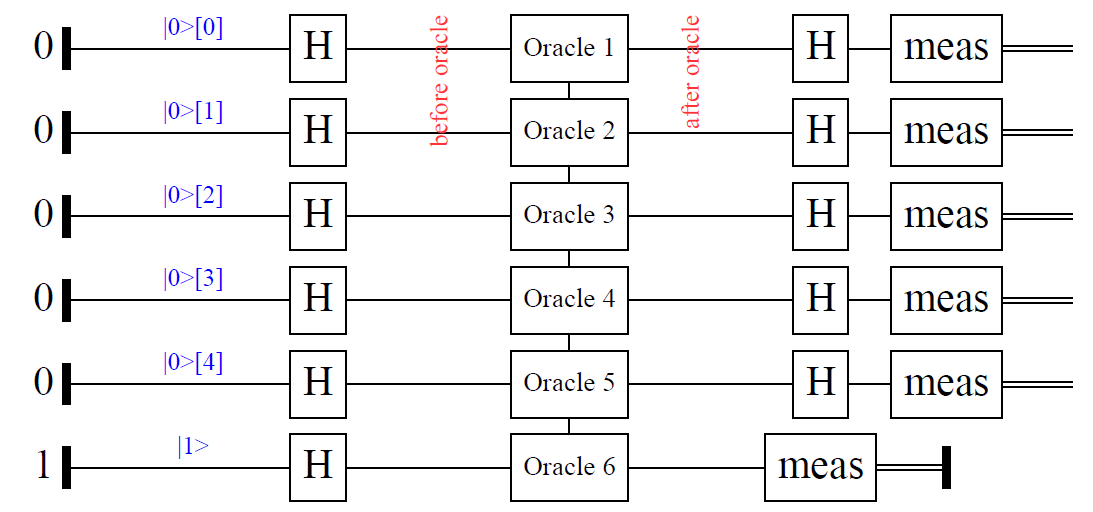}
\caption{Circuit for Deutsch-Jozsa algorithm \cite{noson}}
\end{figure}

\subsubsection{Oracle examples}
\paragraph{Constant Oracle:}
We will write code for constant oracle in \bera{constant\_oracle}. For all inputs of this function, outputs will be either \ensuremath{0} or \ensuremath{1} that means \ensuremath{f(x) = 0} for all \ensuremath{x} or \ensuremath{f(x)=1} for all \ensuremath{x}. The code for \bera{constant\_oracle} is given below:\\

\begin{footnotesize}
\bera{
main = print\_generic Preview (deutsch\_jozsa\_circuit constant\_oracle)
 
~~where 

~~~~-\-- declare constant\_oracle's data type

~~~~constant\_oracle :: Oracle

~~~~constant\_oracle = Oracle \{

~~~~~~qubit\_num = 2,

~~~~~~function = constant\_oracle\_function

~~~~\}

~~~~-\-- initialize constant\_oracle function \ensuremath{f(x)}

~~~~constant\_oracle\_function:: ([Qubit],Qubit) \ensuremath{\rightarrow} Circ ([Qubit],Qubit)

~~~~constant\_oracle\_function (ins,out) = do

~~~~~~-\-- Qubits will remain the same

~~~~~~return (ins, out)
\\}
\end{footnotesize}

\begin{figure}[H]
\centering
\includegraphics[scale=0.4]{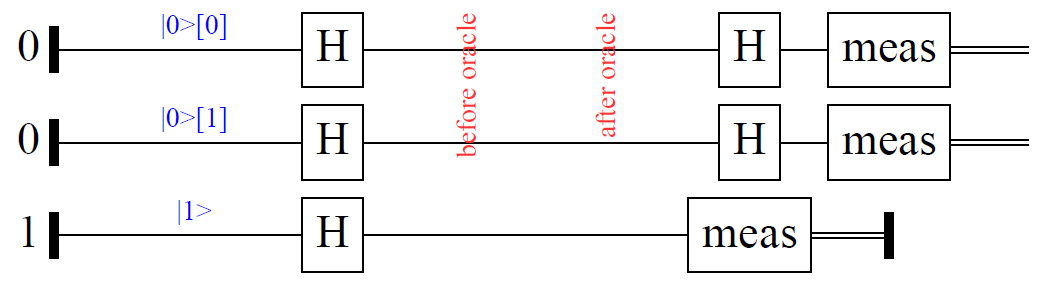}
\caption{Circuit for constant oracle}
\end{figure}

\paragraph{Balanced Oracle:}
We will write code for balanced oracle in \bera{balanced\_oracle}. Exactly half of the inputs for this function will go for \ensuremath{0}'s and other half will go for \ensuremath{1}'s as output that means \ensuremath{f(x)=0} for half of \ensuremath{x} and \ensuremath{f(x)=1} for other half of \ensuremath{x}.\\

\begin{footnotesize}
\bera{
main = print\_generic Preview (deutsch\_jozsa\_circuit balanced\_oracle) 

~~where 

~~~~-\-- declare balanced\_oracle's data type

~~~~balanced\_oracle :: Oracle

~~~~balanced\_oracle = Oracle \{
 
~~~~~qubit\_num = 2,

~~~~~function = balanced\_oracle\_function

~~~~\}

~~~~-\-- initialize balanced\_oracle function \ensuremath{f(x)}

~~~~balanced\_oracle\_function:: ([Qubit],Qubit) \ensuremath{\rightarrow} Circ ([Qubit],Qubit)

~~~~balanced\_oracle\_function ([x,y],out) = do

~~~~~qnot\_at out `controlled` x

~~~~~qnot\_at out `controlled` y

~~~~~return ([x,y],out)

~~~~balanced\_oracle\_function~\_ = error "undefined"  -\-- fallback case
\\}
\end{footnotesize}

\begin{figure}[H]
\centering
\includegraphics[scale=0.4]{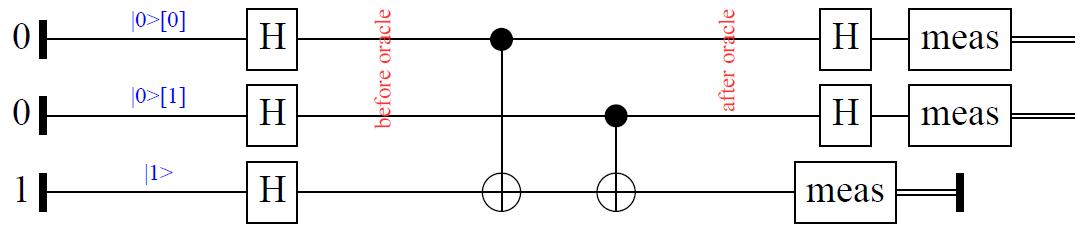}
\caption{Circuit for balanced oracle}
\end{figure}

\subsubsection{Simulation}
Like in the Deutsch algorithm, we will write a new \bera{simulate} function that simulates oracles classically and a new \bera{circuit} function to show the simulation results. \\

\begin{footnotesize}
\bera{
import qualified Data.Map as Map

import QuipperLib.Simulation

import System.Random\\

-\-- simulate function

simulate :: Circ [Bit] \ensuremath{\rightarrow} Bool

simulate oracle = and (map not (run\_generic (mkStdGen 1) (1.0::Float) oracle))
\\}
\end{footnotesize}

Here \bera{simulate} function uses \bera{map} operator to apply \bera{not} to each elements of the oracle to \emph{negate} its values. Then \bera{and} operator is applied to the results and finally returns a \bera{Bool}. \bera{circuit} function will use the \bera{Bool} to print "constant" for constant oracles (when \bera{Bool} is \bera{True}) and "balanced" for balanced oracles (when \bera{Bool} is \bera{False}). Again, \bera{main} function will be used to start the whole program.\\

\begin{footnotesize}
\bera{
-\-- circuit function

circuit :: (Circ [Bit] \ensuremath{\rightarrow} Bool) \ensuremath{\rightarrow} Oracle \ensuremath{\rightarrow} IO ()

circuit run oracle =

~~-\-- first deutsch\_jozsa will apply on oracle

~~-\-- then run function will evaluate the result

~~if run (deutsch\_jozsa\_circuit oracle)
 
~~then putStrLn "constant"

~~else putStrLn "balanced"\\

-\-- main function

main = do

~~-\-- test constant\_oracle

~~circuit simulate constant\_oracle

~~-\-- test balanced\_oracle

~~circuit simulate balanced\_oracle
\\}
\end{footnotesize}

\begin{figure}[H]
\centering
\includegraphics[width=.80\textwidth]{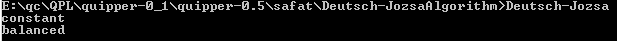}
\caption{Simulation of Deutsch-Jozsa algorithm}
\end{figure}

\subsection{Simon's periodicity algorithm}
Simon's periodicity algorithm is about finding patters in a particular set of functions that's why it is a promised problem. It solves Simon's problem which is a computational problem in the model of decision tree complexity or query complexity, conceived by Daniel Simon in 1994 \cite{simon1994}. It was also the inspiration for Shor's algorithm (we will discuss at section 4.5). Both problems are special cases of the abelian hidden subgroup problem. Simon's algorithm has both quantum procedures and classical procedures. We will mainly discuss about quantum procedures. Simon's algorithm solves the following problem (we quote the definition of the problem verbatim from \cite{noson}):

\begin{myprob}Given a function \ensuremath{f : \{0,1\}^n\rightarrow {\{0,1\}}^n} as a black box, promised to have a secret (hidden) binary string \textbf{s}, such that for all strings \textbf{x, y} \ensuremath{\in {\{0, 1\}}^n}, we have \ensuremath{\textbf{f(x)=f(y)}} if and only if \ensuremath{\textbf{x = y }\oplus\textbf{ s}}. The goal is to determine \textbf{s}.\end{myprob} 

In other words, the values of \ensuremath{f} repeat themselves in some pattern and the pattern is determined by \ensuremath{s}. Function \ensuremath{f} is one to one when \ensuremath{s = 0^n} otherwise \ensuremath{f} is two to one. If we find two inputs \ensuremath{x_1, x_2} such that \ensuremath{f(x_1) = f(x_2)}, then \ensuremath{x_1 = x_2 \oplus s} and we obtain \ensuremath{s} by \ensuremath{x_1 \oplus x_2 = x_2 \oplus s \oplus x_2 = s} 

The worst case scenario for classical algorithm to determine a two to one function is, it needs more than half of the inputs evaluations. So it requires \ensuremath{\frac{2^n}{2}+1 = 2^{n-1} + 1} function evaluations. On the contrary, Simon's algorithm needs \ensuremath{n} function evaluations. Classical computational model needs exponential time complexity to solve this problem while quantum computational model solves it in bounded quantum polynomial time.

\subsubsection{Circuit for Simon's algorithm}
The core section of Simon's algorithm will be written in \bera{simon\_circuit} function. The quantum part of Simon's algorithm is basically performing this function several times. We will write function \bera{steps} for running \bera{simon\_circuit} \ensuremath{n-1} times. This function will return a \bera{Maybe} data type that mean's it may return \bera{([Bit], [Bit])} or may return \bera{Nothing} (readers may think like null value). We will get only the binary string which satisfies \ensuremath{\langle y, s \rangle = 0} \footnote{We are using the standard Dirac notation. \ensuremath{\langle|} is a row vector and \ensuremath{|\rangle} is a column vector. \ensuremath{\langle y, s \rangle} means inner product of \ensuremath{y,s}}. When \ensuremath{\langle y, s \rangle = 1},  destructive interference occurs which cancels each other and we get nothing. Classical part of Simon's algorithm will take that results and solves "linear equations" to find hidden pattern \ensuremath{s}. This can be done by Gaussian elimination, which takes \ensuremath{\Omega{(n^3)}} steps.\\

\begin{footnotesize}
\bera{
import Quipper\\

-\-- declare oracle data type

data Oracle = Oracle \{
    
~~qubit\_num :: Int,

~~function :: ([Qubit], [Qubit]) \ensuremath{\rightarrow} Circ ([Qubit], [Qubit])  

\}\\

-\-- declare simon\_circuit function

simon\_circuit :: Oracle \ensuremath{\rightarrow} Circ ([Bit], [Bit])

simon\_circuit oracle = do

~~-\--create the ancillaes

~~top\_qubits \ensuremath{\leftarrow} qinit (replicate (qubit\_num oracle) False)
                             
~~bottom\_qubits \ensuremath{\leftarrow} qinit (replicate (qubit\_num oracle) True)      
                       
~~label (top\_qubits, bottom\_qubits) ("top |0>", "bottom |1>")

~~-\-- apply first hadamard gate

~~mapUnary hadamard top\_qubits          

~~mapUnary hadamard bottom\_qubits

~~-\-- call the oracle

~~(function oracle) (top\_qubits, bottom\_qubits) 

~~-\-- apply hadamard gate again

~~mapUnary hadamard top\_qubits

~~-\-- measure qubits

~~(top\_qubits, bottom\_qubits) \ensuremath{\leftarrow} measure(top\_qubits, bottom\_qubits)

~~-\-- return the result      

~~return (top\_qubits,bottom\_qubits)\\

-\-- declare steps function

steps :: (Oracle \ensuremath{\rightarrow} Circ ([Bit], [Bit])) \ensuremath{\rightarrow} Oracle \ensuremath{\rightarrow} Circ (Maybe ([Bit], [Bit]))

steps simon\_algorithm oracle = do

~~comment " Simon's algorithm"

~~-\-- set value for n

~~let n = toEnum (qubit\_num oracle) :: Int

~~-\-- call simon\_circuit n-1 times

~~for 1 (n-1) 1 \$ \textbackslash i \ensuremath{\rightarrow} do                    

~~~~comment "start"

~~~~-\-- call simon\_circuit function

~~~~ret  \ensuremath{\leftarrow} simon\_algorithm oracle

~~~~-\-- return the result

~~~~return ret

~~~~comment "finish"

~~endfor     
                      
~~return Nothing\\

-\-- declare main function

main = print\_generic Preview (steps simon\_circuit empty\_oracle)  

~~where 

~~~-\-- declare empty\_oracle's data type

~~~empty\_oracle :: Oracle

~~~empty\_oracle = Oracle \{ 

~~~~~-\-- set the length of qubit string

~~~~~qubit\_num = 4,

~~~~~function = empty\_oracle\_function

~~~\}

~~~-\-- initialize empty\_oracle's function

~~~empty\_oracle\_function:: ([Qubit],[Qubit]) \ensuremath{\rightarrow} Circ ([Qubit],[Qubit])

~~~empty\_oracle\_function (ins,out) = named\_gate "Oracle" (ins,out)
\\}
\end{footnotesize}

\begin{figure}[H]
\centering
\includegraphics[scale=0.6]{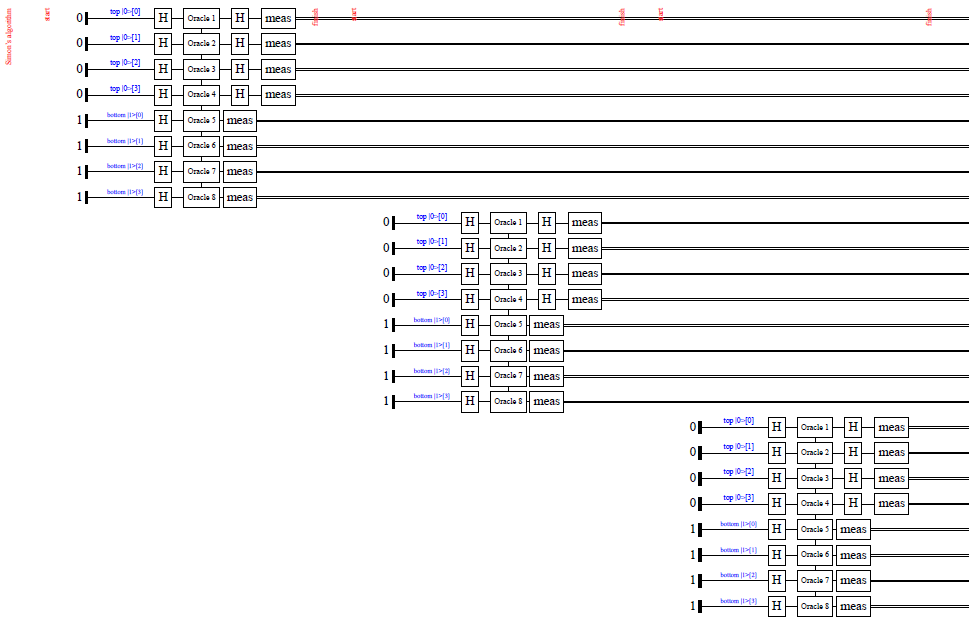}
\caption{Quantum subroutine for Simon's algorithm \cite{s1}}
\end{figure}

\subsubsection{Oracle example}
We will write code for an oracle when \bera{qubit\_num = 2}.  So, the inputs are 00, 01, 10 and 11. Let's consider the period, \textbf{s} = 11. So, we can define the function as follows:\\\\
\ensuremath{f(00) = 01, f(01) = 10, f(10) = f(01 \oplus 11) = 10} and \ensuremath{f(11) = f(00 \oplus 11) = 01}\\\\
The \emph{Unitary Transformation} is: \ensuremath{|x\rangle |y\rangle \rightarrow |x\rangle |y \oplus f(x)\rangle}. We will code the \emph{Unitary matrix} in \bera{sample\_oracle} function. The truth table and the code for \bera{sample\_oracle} function is given below:\\

\begin{table}[ht]
\caption{Truth table of unitary matrix} 
\centering 
\begin{tabular}{c c | c c || c c | c c || c c | c c || c c | c c} 
\hline\hline 
\ensuremath{x} & \ensuremath{y} &\ensuremath{f(x)} & \ensuremath{y\oplus f(x)} & \ensuremath{x} & \ensuremath{y} &\ensuremath{f(x)} & \ensuremath{y\oplus f(x)} & \ensuremath{x} & \ensuremath{y} &\ensuremath{f(x)} & \ensuremath{y\oplus f(x)} & \ensuremath{x} & \ensuremath{y} &\ensuremath{f(x)} & \ensuremath{y\oplus f(x)}\\ [0.5 ex]  
\hline 
00 & 00 & 01 & 01 & 01 & 00 & 10 & 10 & 10 & 00 & 10 & 10 & 11 & 00 & 01 & 01 \\   
00 & 01 & 01 & 00 & 01 & 01 & 10 & 11 & 10 & 01 & 10 & 11 & 11 & 01 & 01 & 00 \\
00 & 10 & 01 & 11 & 01 & 10 & 10 & 00 & 10 & 10 & 10 & 00 & 11 & 10 & 01 & 11 \\  
00 & 11 & 01 & 10 & 01 & 11 & 10 & 01 & 10 & 11 & 10 & 01 & 11 & 11 & 01 & 10 \\  
\hline 
\end{tabular}
\label{table:nonlin} 
\end{table}

\begin{footnotesize}
\bera{
-\-- declare sample\_oracle's data type

sample\_oracle :: Oracle

sample\_oracle = Oracle\{

~~qubit\_num = 2,

~~function = sample\_function  

\}\\

-\-- initialize sample\_oracle's function

sample\_function :: ([Qubit],[Qubit]) \ensuremath{\rightarrow} Circ ([Qubit],[Qubit]) 

sample\_function (controlled\_qubit, target\_qubit) = do

~~let element = controlled\_qubit ++ target\_qubit

~~-\-- call the unitary matrix

~~exact\_synthesis operator element

~~return (controlled\_qubit, target\_qubit)\\

-\-- initialize 16 by 16 unitary matrix

operator :: Matrix Sixteen Sixteen DOmega

operator = matrix16x16 ( 0, 1, 0, 0, 0, 0, 0, 0, 0, 0, 0, 0, 0, 0, 0, 0 )

~~~~~~~~~~~~~~~~~~~~~~~( 1, 0, 0, 0, 0, 0, 0, 0, 0, 0, 0, 0, 0, 0, 0, 0 )

~~~~~~~~~~~~~~~~~~~~~~~( 0, 0, 0, 1, 0, 0, 0, 0, 0, 0, 0, 0, 0, 0, 0, 0 )

~~~~~~~~~~~~~~~~~~~~~~~( 0, 0, 1, 0, 0, 0, 0, 0, 0, 0, 0, 0, 0, 0, 0, 0 )

~~~~~~~~~~~~~~~~~~~~~~~( 0, 0, 0, 0, 0, 0, 1, 0, 0, 0, 0, 0, 0, 0, 0, 0 )

~~~~~~~~~~~~~~~~~~~~~~~( 0, 0, 0, 0, 0, 0, 0, 1, 0, 0, 0, 0, 0, 0, 0, 0 )

~~~~~~~~~~~~~~~~~~~~~~~( 0, 0, 0, 0, 1, 0, 0, 0, 0, 0, 0, 0, 0, 0, 0, 0 )

~~~~~~~~~~~~~~~~~~~~~~~( 0, 0, 0, 0, 0, 1, 0, 0, 0, 0, 0, 0, 0, 0, 0, 0 ) 

~~~~~~~~~~~~~~~~~~~~~~~( 0, 0, 0, 0, 0, 0, 0, 0, 0, 0, 1, 0, 0, 0, 0, 0 )

~~~~~~~~~~~~~~~~~~~~~~~( 0, 0, 0, 0, 0, 0, 0, 0, 0, 0, 0, 1, 0, 0, 0, 0 )

~~~~~~~~~~~~~~~~~~~~~~~( 0, 0, 0, 0, 0, 0, 0, 0, 1, 0, 0, 0, 0, 0, 0, 0 )

~~~~~~~~~~~~~~~~~~~~~~~( 0, 0, 0, 0, 0, 0, 0, 0, 0, 1, 0, 0, 0, 0, 0, 0 )

~~~~~~~~~~~~~~~~~~~~~~~( 0, 0, 0, 0, 0, 0, 0, 0, 0, 0, 0, 0, 0, 1, 0, 0 )

~~~~~~~~~~~~~~~~~~~~~~~( 0, 0, 0, 0, 0, 0, 0, 0, 0, 0, 0, 0, 1, 0, 0, 0 )

~~~~~~~~~~~~~~~~~~~~~~~( 0, 0, 0, 0, 0, 0, 0, 0, 0, 0, 0, 0, 0, 0, 0, 1 )

~~~~~~~~~~~~~~~~~~~~~~~( 0, 0, 0, 0, 0, 0, 0, 0, 0, 0, 0, 0, 0, 0, 1, 0 ) 
\\}
\end{footnotesize}
\begin{figure}[H]
\centering
\includegraphics[scale=0.5]{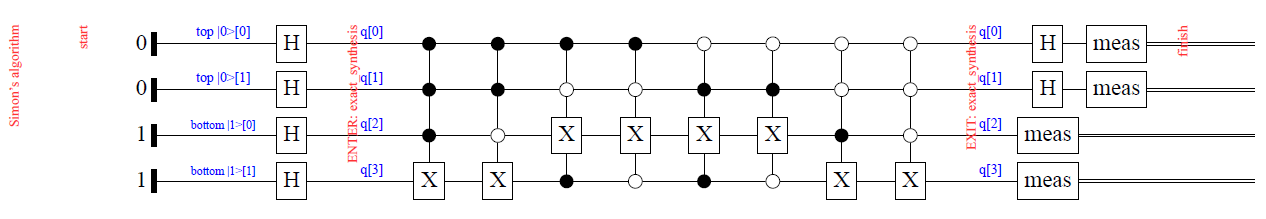}
\caption{Circuit for \bera{sample\_oracle} function}
\end{figure}

\paragraph{16 by 16 matrix constructor :} In previous function, we have implemented \ensuremath{16 \times 16} unitary matrix. As Quipper 0.5 version doesn't include \ensuremath{16 \times 16} matrix constructor, we coded this in \bera{Libraries.Synthesis.Matrix } module. This module provides fixed but arbitrary sized vectors and matrices. Its dimensions are determined by \emph{type level programming}\footnote{calculations are done during compilation time. so it ensures no run-time dimension errors}. We create a type level representation for number \ensuremath{16}. \bera{Ten} and \bera{Succ} are the type representations of number \ensuremath{10} and successor operation, previously declared in this module. \ensuremath{16 \times 16} matrix constructor is \bera{matrix16x16} which takes rows as arguments. Respective codes are given below:\\

\begin{footnotesize}
\bera{
-\-- The natural number 16 as a type

type Sixteen = Succ (Succ (Succ (Succ (Succ (Succ Ten))))) \\

-\-- A convenience constructor for \ensuremath{16 \times 16} matrices

matrix16x16 ::(a, a, a, a, a, a, a, a, a, a, a, a, a, a, a, a)\ensuremath{\rightarrow}

~~~~~~~~~~~~~~(a, a, a, a, a, a, a, a, a, a, a, a, a, a, a, a)\ensuremath{\rightarrow} 

~~~~~~~~~~~~~~(a, a, a, a, a, a, a, a, a, a, a, a, a, a, a, a)\ensuremath{\rightarrow} 

~~~~~~~~~~~~~~(a, a, a, a, a, a, a, a, a, a, a, a, a, a, a, a)\ensuremath{\rightarrow}

~~~~~~~~~~~~~~(a, a, a, a, a, a, a, a, a, a, a, a, a, a, a, a)\ensuremath{\rightarrow}

~~~~~~~~~~~~~~(a, a, a, a, a, a, a, a, a, a, a, a, a, a, a, a)\ensuremath{\rightarrow}

~~~~~~~~~~~~~~(a, a, a, a, a, a, a, a, a, a, a, a, a, a, a, a)\ensuremath{\rightarrow} 

~~~~~~~~~~~~~~(a, a, a, a, a, a, a, a, a, a, a, a, a, a, a, a)\ensuremath{\rightarrow}

~~~~~~~~~~~~~~(a, a, a, a, a, a, a, a, a, a, a, a, a, a, a, a)\ensuremath{\rightarrow}

~~~~~~~~~~~~~~(a, a, a, a, a, a, a, a, a, a, a, a, a, a, a, a)\ensuremath{\rightarrow}

~~~~~~~~~~~~~~(a, a, a, a, a, a, a, a, a, a, a, a, a, a, a, a)\ensuremath{\rightarrow}

~~~~~~~~~~~~~~(a, a, a, a, a, a, a, a, a, a, a, a, a, a, a, a)\ensuremath{\rightarrow}

~~~~~~~~~~~~~~(a, a, a, a, a, a, a, a, a, a, a, a, a, a, a, a)\ensuremath{\rightarrow}

~~~~~~~~~~~~~~(a, a, a, a, a, a, a, a, a, a, a, a, a, a, a, a)\ensuremath{\rightarrow}

~~~~~~~~~~~~~~(a, a, a, a, a, a, a, a, a, a, a, a, a, a, a, a)\ensuremath{\rightarrow}

~~~~~~~~~~~~~~(a, a, a, a, a, a, a, a, a, a, a, a, a, a, a, a)\ensuremath{\rightarrow}Matrix Sixteen Sixteen~a\\

matrix16x16~~~(a0, a1, a2, a3, a4, a5, a6, a7, a8, a9, a10, a11, a12, a13, a14, a15)

~~~~~~~~~~~~~~(b0, b1, b2, b3, b4, b5, b6, b7, b8, b9, b10, b11, b12, b13, b14, b15) 

~~~~~~~~~~~~~~(c0, c1, c2, c3, c4, c5, c6, c7, c8, c9, c10, c11, c12, c13, c14, c15) 

~~~~~~~~~~~~~~(d0, d1, d2, d3, d4, d5, d6, d7, d8, d9, d10, d11, d12, d13, d14, d15)

~~~~~~~~~~~~~~(e0, e1, e2, e3, e4, e5, e6, e7, e8, e9, e10, e11, e12, e13, e14, e15) 

~~~~~~~~~~~~~~(f0, f1, f2, f3, f4, f5, f6, f7, f8, f9, f10, f11, f12, f13, f14, f15) 

~~~~~~~~~~~~~~(g0, g1, g2, g3, g4, g5, g6, g7, g8, g9, g10, g11, g12, g13, g14, g15)

~~~~~~~~~~~~~~(h0, h1, h2, h3, h4, h5, h6, h7, h8, h9, h10, h11, h12, h13, h14, h15)

~~~~~~~~~~~~~~(i0, i1, i2, i3, i4, i5, i6, i7, i8, i9, i10, i11, i12, i13, i14, i15) 

~~~~~~~~~~~~~~(j0, j1, j2, j3, j4, j5, j6, j7, j8, j9, j10, j11, j12, j13, j14, j15) 

~~~~~~~~~~~~~~(k0, k1, k2, k3, k4, k5, k6, k7, k8, k9, k10, k11, k12, k13, k14, k15) 

~~~~~~~~~~~~~~(l0, l1, l2, l3, l4, l5, l6, l7, l8, l9, l10, l11, l12, l13, l14, l15)

~~~~~~~~~~~~~~(m0, m1, m2, m3, m4, m5, m6, m7, m8, m9, m10, m11, m12, m13, m14, m15) 

~~~~~~~~~~~~~~(n0, n1, n2, n3, n4, n5, n6, n7, n8, n9, n10, n11, n12, n13, n14, n15) 

~~~~~~~~~~~~~~(o0, o1, o2, o3, o4, o5, o6, o7, o8, o9, o10, o11, o12, o13, o14, o15) 

~~~~~~~~~~~~~~(p0, p1, p2, p3, p4, p5, p6, p7, p8, p9, p10, p11, p12, p13, p14, p15)~= \\

matrix~~~~~~~~[[a0, b0, c0, d0, e0, f0, g0, h0, i0, j0, k0, l0, m0, n0, o0, p0], 

~~~~~~~~~~~~~~[a1, b1, c1, d1, e1, f1, g1, h1, i1, j1, k1, l1, m1, n1, o1, p1], 

~~~~~~~~~~~~~~[a2, b2, c2, d2, e2, f2, g2, h2, i2, j2, k2, l2, m2, n2, o2, p2], 

~~~~~~~~~~~~~~[a3, b3, c3, d3, e3, f3, g3, h3, i3, j3, k3, l3, m3, n3, o3, p3],

~~~~~~~~~~~~~~[a4, b4, c4, d4, e4, f4, g4, h4, i4, j4, k4, l4, m4, n4, o4, p4], 

~~~~~~~~~~~~~~[a5, b5, c5, d5, e5, f5, g5, h5, i5, j5, k5, l5, m5, n5, o5, p5], 

~~~~~~~~~~~~~~[a6, b6, c6, d6, e6, f6, g6, h6, i6, j6, k6, l6, m6, n6, o6, p6], 

~~~~~~~~~~~~~~[a7, b7, c7, d7, e7, f7, g7, h7, i7, j7, k7, l7, m7, n7, o7, p7], 

~~~~~~~~~~~~~~[a8, b8, c8, d8, e8, f8, g8, h8, i8, j8, k8, l8, m8, n8, o8, p8], 

~~~~~~~~~~~~~~[a9, b9, c9, d9, e9, f9, g9, h9, i9, j9, k9, l9, m9, n9, o9, p9], 

~~~~~~~~~~~~~~[a10,b10,c10,d10,e10,f10,g10,h10,i10,j10,k10,l10,m10,n10,o10,p10], 

~~~~~~~~~~~~~~[a11,b11,c11,d11,e11,f11,g11,h11,i11,j11,k11,l11,m11,n11,o11,p11], 

~~~~~~~~~~~~~~[a12,b12,c12,d12,e12,f12,g12,h12,i12,j12,k12,l12,m12,n12,o12,p12], 

~~~~~~~~~~~~~~[a13,b13,c13,d13,e13,f13,g13,h13,i13,j13,k13,l13,m13,n13,o13,p13], 

~~~~~~~~~~~~~~[a14,b14,c14,d14,e14,f14,g14,h14,i14,j14,k14,l14,m14,n14,o14,p14], 

~~~~~~~~~~~~~~[a15,b15,c15,d15,e15,f15,g15,h15,i15,j15,k15,l15,m15,n15,o15,p15]] 
}
\end{footnotesize}
\paragraph{For 8 by 8 matrix:} We also write code for \ensuremath{8 \times 8} matrix constructor named \bera{matrix8x8} in \bera{Libraries.Synthesis.Matrix} module. It also takes rows as argument.\\

\begin{footnotesize}
\bera{
-\--  define \ensuremath{8 \times 8} matrix constructor

matrix8x8::(a, a, a, a, a, a, a, a) \ensuremath{\rightarrow} (a, a, a, a, a, a, a, a) \ensuremath{\rightarrow} 

~~~~~~~~~~~(a, a, a, a, a, a, a, a) \ensuremath{\rightarrow} (a, a, a, a, a, a, a, a) \ensuremath{\rightarrow}

~~~~~~~~~~~(a, a, a, a, a, a, a, a) \ensuremath{\rightarrow} (a, a, a, a, a, a, a, a) \ensuremath{\rightarrow} 

~~~~~~~~~~~(a, a, a, a, a, a, a, a) \ensuremath{\rightarrow} (a, a, a, a, a, a, a, a) \ensuremath{\rightarrow} Matrix Eight Eight~a\\

matrix8x8~~(a0, a1, a2, a3, a4, a5, a6, a7) (b0, b1, b2, b3, b4, b5, b6, b7) 

~~~~~~~~~~~(c0, c1, c2, c3, c4, c5, c6, c7) (d0, d1, d2, d3, d4, d5, d6, d7)

~~~~~~~~~~~(e0, e1, e2, e3, e4, e5, e6, e7) (f0, f1, f2, f3, f4, f5, f6, f7) 

~~~~~~~~~~~(g0, g1, g2, g3, g4, g5, g6, g7) (h0, h1, h2, h3, h4, h5, h6, h7)~= \\

matrix~~~~[[a0, b0, c0, d0, e0, f0, g0, h0], [a1, b1, c1, d1, e1, f1, g1, h1], 

~~~~~~~~~~~[a2, b2, c2, d2, e2, f2, g2, h2], [a3, b3, c3, d3, e3, f3, g3, h3], 

~~~~~~~~~~~[a4, b4, c4, d4, e4, f4, g4, h4], [a5, b5, c5, d5, e5, f5, g5, h5], 

~~~~~~~~~~~[a6, b6, c6, d6, e6, f6, g6, h6], [a7, b7, c7, d7, e7, f7, g7, h7]]
}
\end{footnotesize}

\subsection{Grover's search algorithm}
Grover's search algorithm is a quantum algorithm for searching an unsorted database with \ensuremath{N} entries in O(\ensuremath{\sqrt{N}}) time and using O(\ensuremath{\log N}) storage space, invented by Lov Grover in 1996 \cite{grover1996fast}. Grover's search algorithm solves the following problem (we quote the definition of the problem verbatim from \cite{noson}):

\begin{myprob}Given a function \ensuremath{f : \{0,1\}^n\rightarrow \{0,1\}} as a black box, exists exactly one binary string \ensuremath{x_0} such that \ensuremath{f(x) = 1} if \ensuremath{x = x_0} and \ensuremath{f(x) = 0} if \ensuremath{x \neq x_0}. The goal is to find \ensuremath{x_0}.\end{myprob} 

Classically to solve this problem, it needs \ensuremath{\frac{N}{2}} time on average and \ensuremath{N} time in worst case scenario. Unlike previous quantum algorithms, Grover's search algorithm provides a quadratic speedup. Some of the classical algorithms have linear time complexity and quantum algorithm solves the same problem in complexity class BQP.

It uses two tricks to increase the probability of desired binary string \ensuremath{x_0}.
\begin{enumerate} 
\item \textbf{Phase inversion}: is used to change the phase of the desired state.
\item \textbf{Inversion about the average}: is used to boost the separation of the phases.
\end{enumerate}
These operations should not be done more than \ensuremath{\sqrt{N}} times. Otherwise, for over computation the probability of desired binary string may decrease.
\subsubsection{Circuit for Grover's search algorithm}
The core section of Grover's search algorithm is in \bera{grover\_search\_circuit} function. This function calls \bera{phase\_inversion} and \bera{inversion\_about\_mean} functions for \ensuremath{\sqrt{2^n}} times. These functions act as their names suggest. \bera{phase\_inversion} function will take an oracle, a qubit string and apply that oracle function on that qubit string. \bera{inversion\_about\_mean} function will separate target qubit and controlled qubit from qubit string and apply \ensuremath{2 |\psi\rangle \langle\psi| - I}, the conditional phase shift operation \cite{g3}. Finally \bera{main} function will call \bera{grover\_search\_circuit} with a dummy oracle named \bera{empty\_oracle}. Respective codes are given below:\\

\begin{footnotesize}
\bera{
import Quipper\\

-\-- declare Oracle data type

data Oracle = Oracle \{
      
~~qubit\_num :: Int,

~~function :: ([Qubit], Qubit) \ensuremath{\rightarrow} Circ ([Qubit], Qubit)  

\}\\

-\-- declare phase\_inversion function

phase\_inversion::(([Qubit],Qubit)\ensuremath{\rightarrow}Circ([Qubit],Qubit))\ensuremath{\rightarrow}([Qubit],Qubit)\ensuremath{\rightarrow} Circ([Qubit],Qubit)

phase\_inversion oracle (top\_qubits, bottom\_qubit) = do

~~comment "start phase inversion"      

~~-\-- call oracle                    

~~oracle (top\_qubits, bottom\_qubit)                       

~~comment "end phase inversion"

~~return (top\_qubits, bottom\_qubit)\\

-\-- declare inversion\_about\_mean function

inversion\_about\_mean :: ([Qubit], Qubit) \ensuremath{\rightarrow} Circ ([Qubit], Qubit)

inversion\_about\_mean (top\_qubits, bottom\_qubit) = do

~~comment "start inversion about mean"

~~-\-- apply X gate at top qubit

~~mapUnary gate\_X top\_qubits

~~-\-- separate target and control qubits

~~let pos = (length top\_qubits) - 1

~~let target\_qubit = top\_qubits !! pos

~~let controlled\_qubit = take pos top\_qubits
   
~~-\-- apply hadamard at target\_qubit

~~hadamard\_at target\_qubit

~~-\-- apply qnot gate at target qubit

~~qnot\_at target\_qubit `controlled` controlled\_qubit
  
~~-\-- apply hadamard again at top
  
~~hadamard\_at target\_qubit

~~-\-- apply X gate at bottom

~~mapUnary gate\_X top\_qubits
 
~~comment "end inversion about mean"

~~return (top\_qubits, bottom\_qubit)\\

-\-- declare grover\_search\_circuit function

grover\_search\_circuit :: Oracle \ensuremath{\rightarrow} Circ ([Bit])

grover\_search\_circuit oracle = do

~~comment "Grover Search algorithm"

~~-\-- set the value of n

~~let n = toEnum (qubit\_num oracle) :: Float

~~-\-- set the index number to iterate \ensuremath{\sqrt{2^n}} times

~~let index = (floor (sqrt (2**n))) :: Int

~~-\-- create the ancillaes

~~top \ensuremath{\leftarrow} qinit (replicate (qubit\_num oracle) False)       

~~bottom \ensuremath{\leftarrow} qinit True  

~~label (top, bottom) ("|0>","|1>")

~~-\-- apply hadamard gate at string

~~mapUnary hadamard top

~~hadamard\_at bottom

~~-\-- start to iterate

~~for 1 (index) 1 \$ \textbackslash i \ensuremath{\rightarrow} do                                     

~~~~~comment "start grover iteration" 

~~~~~-\-- call phase inversion 

~~~~~(top, bottom) \ensuremath{\leftarrow} phase\_inversion (function oracle) (top, bottom) 

~~~~~-\-- call inversion about mean

~~~~~(top, bottom) \ensuremath{\leftarrow} inversion\_about\_mean (top, bottom)

~~~~~comment "after grover iteration"

~~endfor

~~-\-- measure qubit string and return result

~~hadamard\_at bottom

~~(top, bottom) \ensuremath{\leftarrow} measure (top, bottom)                                 

~~cdiscard bottom

~~return (top)\\

-\-- main function

main = print\_generic Preview (grover\_search\_circuit empty\_oracle)

~~where

~~~-\-- declare empty\_oracle's data type

~~~empty\_oracle :: Oracle

~~~empty\_oracle = Oracle \{ 

~~~~~-\-- set the length of qubit string

~~~~~qubit\_num = 4,

~~~~~function = empty\_oracle\_function

~~~\}

~~~-\-- initialize empty\_oracle's function \ensuremath{f(x)}

~~~empty\_oracle\_function:: ([Qubit],Qubit) \ensuremath{\rightarrow} Circ ([Qubit],Qubit)

~~~empty\_oracle\_function (ins,out) = named\_gate "Oracle" (ins,out)
\\}
\end{footnotesize}

\begin{figure}[H]
\centering
\includegraphics[width=\textwidth]{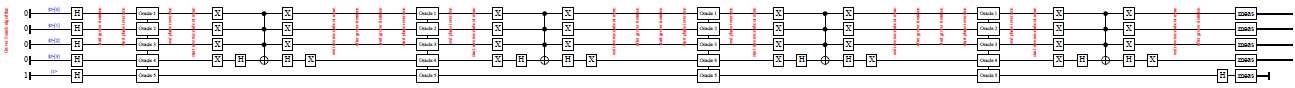}
\caption{Circuit for Grover's search algorithm}
\end{figure}

\subsubsection{Oracle examples}

\paragraph{Single Grover iteration :} We will code for single Grover iteration in \bera{oracle\_two}. This oracle will find the search element \ensuremath{x_0=2} in search space size \ensuremath{N=4} \cite{g3}. So oracle's reaction will be \ensuremath{f(x) = 1} when \ensuremath{x=x_0=2} and \ensuremath{f(x)=0} when \ensuremath{x \neq 2}. \bera{oracle\_two} function is:\\

\begin{footnotesize}
\bera{
-\-- declare oracle\_two data type

oracle\_two :: Oracle

oracle\_two = Oracle \{

~~qubit\_num = 2,

~~function = oracle\_two\_function

\}

-\-- initialize oracle\_two function \ensuremath{f(x)}

oracle\_two\_function :: ([Qubit],Qubit) \ensuremath{\rightarrow} Circ ([Qubit],Qubit)

oracle\_two\_function (controlled\_qubit, target\_qubit) = do

~~qnot\_at target\_qubit `controlled` controlled\_qubit .==. [1,0]

~~return (controlled\_qubit, target\_qubit) 
\\}
\end{footnotesize}

\begin{figure}[H]
\centering
\includegraphics[scale=0.5]{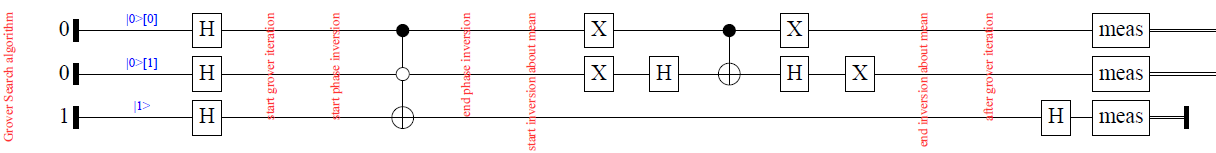}
\caption{Circuit for Grover's single iteration \ensuremath{(x_0=2)} \cite{g2} \cite{g3}}
\end{figure}

\paragraph{Multiple Grover iteration :} \bera{oracle\_five} function needs multiple Grover's iteration to fine search element \ensuremath{x_0=5} in search space size \ensuremath{N=8}. This function will behave like \ensuremath{f(x) = 1} when \ensuremath{x = x_0 = 5} and \ensuremath{f(x)=0} when \ensuremath{x \neq 5}. \bera{oracle\_five} function is:\\

\begin{footnotesize}
\bera{
-\-- declare oracle\_five's data type

oracle\_five :: Oracle

oracle\_five = Oracle \{

~~qubit\_num = 3,

~~function = oracle\_five\_function

\}

-\-- initialize oracle\_five's function

oracle\_five\_function :: ([Qubit],Qubit) \ensuremath{\rightarrow} Circ ([Qubit],Qubit)

oracle\_five\_function (controlled\_qubit, target\_qubit) = do

~~qnot\_at target\_qubit `controlled` controlled\_qubit .==. [1,0,1]

~~return (controlled\_qubit, target\_qubit) 
\\}
\end{footnotesize}

\begin{figure}[H]
\centering
\includegraphics[scale=0.4]{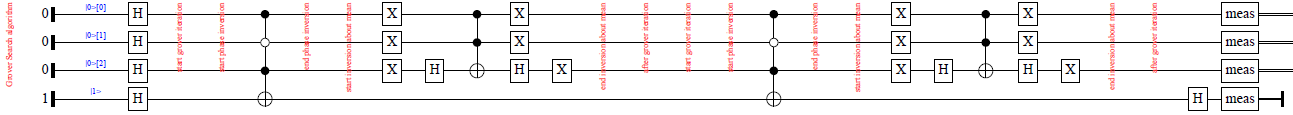}
\caption{Circuit for Grover's multiple iteration \ensuremath{(x_0=5)} \cite{g2}}
\end{figure}

\subsection{Shor's factoring algorithm}
Shor's algorithm, the most celebrated quantum algorithm is formulated by Peter Shor in 1994 \cite{shor1994algorithms} for integer factorization. Like Simon's periodicity algorithm, Shor' algorithm uses phase estimation procedure to factor integers in polynomial time. This factoring problem turns out to be equivalent to the order finding problem discussed in Simon's algorithm. Shor's algorithm uses two basic steps to reduce factoring procedure to order finding procedure. These two steps are embodied in the following theorems (we quote these theorems verbatim from \cite{g3}):

\begin{thm}Suppose N is a n bit composite number, and \textbf{a} is a non-trivial solution to the equation \ensuremath{\textbf{a}^2} = 1(mod N) in the range 1\ensuremath{\le} \textbf{a} \ensuremath{\le} N, that is, neither \textbf{a} = 1(mod~N) nor \textbf{a} = N\ensuremath{- 1} = \ensuremath{- 1}(mod~N). Then at least one of gcd(\ensuremath{\textbf{a}-1}, N) and gcd(\ensuremath{\textbf{a} + 1}, N) is a non-trivial factor of N.\end{thm}

\begin{thm}Suppose N = \ensuremath{p_1^{\alpha_1} . . p_m^{\alpha_m}} is the prime factorization of an odd composite positive integer. Let \textbf{a} be an integer chosen uniformly at random, subject to the requirements that \ensuremath{1\le \textbf{a} \le N - 1} and \textbf{a} is co-prime to N. Let r be the order of \textbf{a} modulo N. Then\\ \\ p(r is even and \ensuremath{\textbf{a}^{r/2} \neq -1 (mod~N)) \ge 1 - \ensuremath{\frac{1}{2^m}}}\end{thm}

All these theorems can be implemented in classical part, but quantum part needs to find the periods of the functions that hold big integers. We can formally state Shor's algorithm like below \cite{noson}:
\paragraph{Step 1:}First check if \ensuremath{N} is a prime or a power of prime. Polynomial algorithm can be used to determine it. If yes then declare it and exit. Otherwise go to the next step.
\paragraph{Step 2:}Choose a random integer \ensuremath{\textbf{a}} such that \ensuremath{1< \textbf{a} < N} and then determine \ensuremath{gcd(\textbf{a}, N)} using Euclid's algorithm. If the GCD is not 1, then return it and exit.
\paragraph{Step 3:}Quantum subroutine for finding the period \ensuremath{r} of a function.
\paragraph{Step 4:}If \ensuremath{r} is odd or if \ensuremath{\textbf{a}^r \equiv -1~mod~N}, then chosen \ensuremath{\textbf{a}} is not appropriate for further calculations. Go back to step 2 and choose another \ensuremath{\textbf{a}}.
\paragraph{Step 5:}Calculate \ensuremath{gcd((\textbf{a}^{\frac{r}{2}}+1),N)} and \ensuremath{gcd((\textbf{a}^{\frac{r}{2}}-1),N)} using Euclid's algorithm and return at least one of the nontrivial solutions.\\

Shor's algorithm needs \ensuremath{O(n^2~log~n~log~log~n)} number of steps, where \ensuremath{n} is the number of bit to represent \ensuremath{N}. On the contrary, best known classical algorithm requires \ensuremath{O(\exp ^{cn^{\frac{1}{3}} log^{\frac{2}{3}}n})} where \ensuremath{c} is some constant. Classical algorithms work in sub-exponential time and Shor's algorithm solves it in the complexity class BQP. Here quantum algorithm gives an exponential speedup. Here we will mainly discuss the quantum subroutine of Shor's algorithm.

\subsubsection{Circuit for Shor's algorithm}
The main section of Shor's algorithm is written in \bera{shor\_circuit} function. In this function, oracle \ensuremath{f_{(a,N)}(x)} will be applied to find it's period \ensuremath{r}. Then \ensuremath{QFT^\dagger} \footnote{adjoint matrix of \ensuremath{QFT}} will be applied to modify the period \ensuremath{r} into \ensuremath{\frac{2^{top\_num}}{r}} and to eliminate the offset. Quipper provides \bera{QuipperLib.QFT} module to perform \emph{Quantum Fourier Transformation}. We will use \bera{qft\_big\_endian} function of this module to apply \ensuremath{QFT^\dagger} on \bera{top\_qubit}. Finally \bera{main} function will pass a dummy oracle named \bera{empty\_oracle} to \bera{shor\_circuit} function and start the whole program.\\

\begin{footnotesize}
\bera{
import Quipper

import QuipperLib.QFT\\

-\-- define oracle data type

data Oracle = Oracle \{

~~top\_num :: Int,

~~bottom\_num :: Int,

~~function :: ([Qubit], [Qubit]) \ensuremath{\rightarrow} Circ ([Qubit], [Qubit])  

\}\\

-\-- declare shor\_circuit function

shor\_circuit :: Oracle \ensuremath{\rightarrow} Circ [Bit]

shor\_circuit oracle = do

~~comment "Shor algorithm"

~~-\-- create the ancillaes

~~top\_qubit \ensuremath{\leftarrow} qinit (replicate (top\_num oracle) False)                             

~~bottom\_qubit \ensuremath{\leftarrow} qinit (replicate (bottom\_num oracle) False)                       

~~label (top\_qubit, bottom\_qubit) ("top\_qubit", "bottom\_qubit")

~~-\-- apply hadamard at top qubits

~~mapUnary hadamard top\_qubit  

~~comment "applying oracle"

~~-\-- call the oracle

~~function oracle (top\_qubit, bottom\_qubit) 

~~comment "after oracle"

~~-\-- measure bottom qubits and discard

~~bottom\_qubit  \ensuremath{\leftarrow} measure bottom\_qubit 

~~cdiscard bottom\_qubit

~~-\-- apply \ensuremath{QFT^{\dagger}}

~~top\_qubit  \ensuremath{\leftarrow} qft\_big\_endian top\_qubit

~~-\-- measure top qubits and return results

~~top\_qubit  \ensuremath{\leftarrow} measure top\_qubit

~~return top\_qubit\\

-\-- main function

main = print\_generic Preview (shor\_circuit empty\_oracle) 

~~where   

~~~-\-- declare empty\_oracle's data type

~~~empty\_oracle :: Oracle

~~~empty\_oracle = Oracle \{

~~~~~top\_num = 6,

~~~~~bottom\_num = 3,

~~~~~function = empty\_function  

~~~\}

~~~-\-- initialize empty\_oracle's function \ensuremath{f(x)}

~~~empty\_function :: ([Qubit], [Qubit]) \ensuremath{\rightarrow} Circ ([Qubit], [Qubit]) 

~~~empty\_function (top, bottom) = named\_gate "Oracle" (top,bottom)
\\}
\end{footnotesize}

\begin{figure}[H]
\centering
\includegraphics[width=\textwidth]{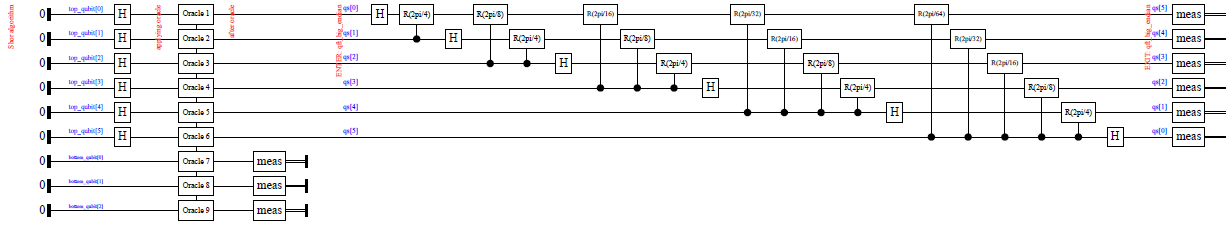}
\caption{Quantum subroutine for Shor's algorithm\cite{shor3}}
\end{figure}

\subsubsection{Oracle examples}
We will write an oracle \ensuremath{f_{(a,N)}(x)} where \ensuremath{a = 7} and \ensuremath{N = 15} \cite{shor3}. Corresponding oracle \bera{mod15\_base7\_oracle} is given below:\\

\begin{footnotesize}
\bera{
-\-- declare mod15\_base7\_oracle data type

mod15\_base7\_oracle :: Oracle

mod15\_base7\_oracle = Oracle\{

~~top\_num = 3,

~~bottom\_num = 4,

~~function = mod15\_base7\_function  

\}\\

-\-- initialize mod15\_base7\_oracle's function \ensuremath{f(x)}

mod15\_base7\_function :: ([Qubit], [Qubit]) \ensuremath{\rightarrow} Circ ([Qubit], [Qubit]) 

mod15\_base7\_function (top\_qubit, bottom\_qubit) = do

~~let x1 = top\_qubit !! 1

~~let x2 = top\_qubit !! 2

~~let y0 = bottom\_qubit !! 0

~~let y1 = bottom\_qubit !! 1

~~let y2 = bottom\_qubit !! 2

~~let y3 = bottom\_qubit !! 3

~~qnot\_at y1 `controlled` x2

~~qnot\_at y2 `controlled` x2

~~qnot\_at y2 `controlled` y0

~~qnot\_at y0 `controlled` [x1,y2]

~~qnot\_at y2 `controlled` y0

~~qnot\_at y1 `controlled` y3

~~qnot\_at y3 `controlled` [x1,y1]

~~qnot\_at y1 `controlled` y3

~~return (top\_qubit, bottom\_qubit)
\\}
\end{footnotesize}

\begin{figure}[H]
\centering
\includegraphics[scale=0.5]{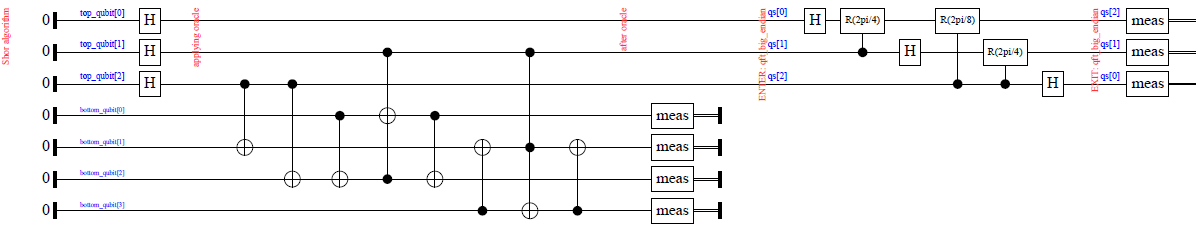}
\caption{Circuit for an oracle of \ensuremath{f_{(a,N)}, a=7 , N = 15} \cite{shor3}}
\end{figure}

We will write another oracle \ensuremath{f_{(a,N)}(x)} where \ensuremath{a = 20} and \ensuremath{N = 21}. We will optimize quantum circuits for modular multiplication \cite{shor2}. Code for \bera{mod21\_base20\_oracle} is given below:\\

\begin{footnotesize}
\bera{
-\-- declare mod21\_base20\_oracle data type

mod21\_base20\_oracle :: Oracle

mod21\_base20\_oracle = Oracle\{

~~top\_num = 3,

~~bottom\_num = 5,

~~function = mod21\_base20\_function  

\}

-\-- initialize mod21\_base20\_oracle function

mod21\_base20\_function :: ([Qubit], [Qubit]) \ensuremath{\rightarrow} Circ ([Qubit], [Qubit]) 

mod21\_base20\_function (top\_qubit, bottom\_qubit) = do

~~-\-- separate control qubit

~~let cntrl\_qbit = head top\_qubit

~~-- separate \ensuremath{y_0, y_2} and \ensuremath{y_4}

~~let y0 = head bottom\_qubit

~~let y2 = head (drop 2 bottom\_qubit)

~~let y4 = head (drop 4 bottom\_qubit)

~~-\-- apply quantum gates

~~qnot\_at y4 `controlled` cntrl\_qbit

~~qnot\_at y2 `controlled` cntrl\_qbit 

~~qnot\_at y0 `controlled` cntrl\_qbit .==. 0        

~~return (top\_qubit, bottom\_qubit)
\\}
\end{footnotesize}

\begin{figure}[H]
\centering
\includegraphics[scale=0.5]{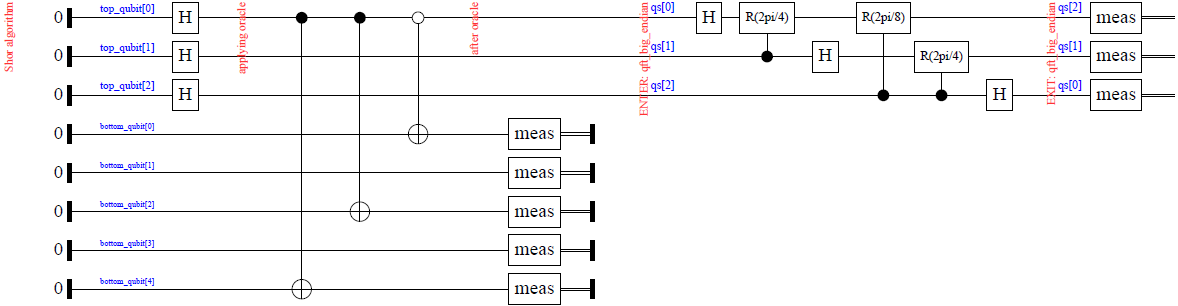}
\caption{Circuit for an oracle of \ensuremath{f_{(a,N)}, a=20 , N = 21} \cite{shor2}}
\end{figure}

%
%
%
%
%
%
%
%
%
%
%
%
%
%
%
%
%
%
%
%
%
%
%
%
%
%
%
%
%
%
%
%
%
%

\section{Conclusion}
This report mainly targets the readers who are interested to study functional quantum programming language.  We briefly discussed about some other classes of quantum programming languages. We have introduced readers to Quipper and implemented five quantum algorithms using it. It demonstrates the capability of the language and the work overhead for quantum programmers. Starting from a "hello world" program we demonstrated up to Shor's factoring algorithm in a pedagogically suitable sequence in order to ensure a smooth learning carve. We have also given examples of high dimensional oracles (black boxes or functions) for specific algorithms and test some of them using Quipper simulator.
\section{Acknowledgement}
SS thanks the Quipper authors  Beno\^it Valiron and Peter Selinger for their continuous help in understanding the language throughout the project. He also likes to thank Professor Md. Shahidur Rahman for giving permission to pursue this project through an experimental distance learning setup. OS thanks Professor Samuel J. Lomonaco Jr. for his comments.
\bibliography{paper}
\bibliographystyle{plain}
\end{small}
\end{document}